# Thermodynamics of data

Borko Stosic

Universidade Federal Rural de Pernambuco, Rua Dom Manoel de Medeiros s/n, 52171-900 - Dois Irmãos, Recife-PE, Brazil

## Abstract

The recently introduced concept of generalized thermodynamics is explored here in the context of 1d, 2d and 3d data analysis, performed on samples drawn from a 3d X-ray soil sample image. Different threshold levels are used to binarize the 3d sample, wherefrom relative frequencies of binary patterns are found and then used to address finite size scaling behavior of the response functions as a function of the disorder parameter (equivalent of temperature in thermodynamics). It is found that for different threshold levels response functions for increasing sample sizes approach the thermodynamic limit from different directions, with a crossover reminiscent of a transition from open to periodic boundaries of the Ising model, implying existence of a characteristic correlation scale. It is argued here that this characteristic scale corresponds to the "natural" properties of the data, where correlations within finite size samples are neither underestimated nor overestimated. In the current context of soil this scale may be related to the so-called Representative elementary volume (REV), while in other situations this characteristic scale should be interpreted in the context of the phenomenon under study.

## Introduction

It was shown a decade ago [1,2] how the formidable formalism of statistical mechanics can be employed to deal with empirical frequency distributions, offering a new perspective that may prove useful for data analysis in general. In [1] this concept was introduced to study statistical thermodynamics of natural images, and in [2] it was used to address organization of resting brain activity. The theoretical basis of statistical thermodynamics of probability distributions was addressed some years later [3], providing solid grounds for the previous (as well as future) works, but it seems that up to date it has not attracted due attention from the scientific community. In short, statistical mechanics formalism (that was introduced to make sense of phenomenological thermodynamics observations) is not restricted only to systems of interacting particles – it can be generalized to enhance our understanding of empirical data from diverse phenomena in general.

On the other hand, in Refs. [1,2] the data are analyzed using a single (arbitrary) threshold. In the current work the data is analyzed for a range of different threshold values, revealing the crossover between the range of values where correlations are overestimated for finite size samples, and the region where correlations are underestimated. This particular characteristic threshold where this transition occurs corresponds to "scale free" behavior, where the sample size ceases to affect the position of maxima of the finite size sample response functions (specific heat and susceptibility).

As the computational power and data acquisition capabilities have been growing at an exponential rate over the last decades, the concepts of "small", "large", "reasonable" and "beyond reach" sample size have been changing in parallel. The practical questions that we could not even pose some decades back when a sample of size $n = 30$ was regarded "reasonable" and that of $n = 100$ was regarded "large" are now becoming amenable. It seems paramount that we adjust our data analysis methods to these changes, and the current approach of employing statistical thermodynamics of probability distributions for data analysis seems to be in the right direction.

As a working example 3d X-ray CT scan data are used in this work to extract 1d, 2d and 3d samples to address the question of Representative elementary volume (REV). The concept of REVs has been a subject of much recent research (see e.g. [4,5] and references therein), with no clear consensus on the scientific community's REV definition in sight. In the direction of elucidating the REV concept, in this work the generalized thermodynamics formalism [1,2] is employed to analyze the finite size scaling properties of a 3d X-ray soil sample, leading to the conjecture that REV could be associated with scale free response function finite size sample behavior. In particular, it is found that for all three sampling schemes (1d, 2d and 3d) the response functions for increasing sample sizes below and above the same threshold level approach the thermodynamic limit from different directions, representing a crossover reminiscent of a transition from open to periodic boundaries of the Ising model, where paths connecting point pairs of finite size systems are underestimated for the former, and overestimated for the latter. At the crossover threshold the response functions are scale free, and the clusters observed at this threshold may be viewed as characteristic for the sample under study.

This novel approach may lead to new practical progress in a wide range of disciplines where data analysis is of fundamental importance.

## Methodology

### *Data*

The soil sample used in this study was collected from a sugar cane field located in the state of Pernambuco, in the northeastern region of Brazil, as part of previous studies [6–8]. The soil at the sampling point was classified as a Latossolo in the Brazilian soil classification system, equivalent to Ferralsol in the World Reference Base for Soil Resources (WRB) or Oxisol in the United States Soil Taxonomy System, collected from the 0-10 cm soil layer. The CT tomography of the soil samples was performed at the X-ray Computed Tomography Laboratory of the Nuclear Energy Department at UFPE - Federal University of Pernambuco, Brazil. Details of the data acquisition procedure are presented in [8], resulting in $790^3$=493,039,000 voxels, at 45μm spatial resolution, with values in Hounsfield units (HU).

### *Thermodynamics of data*

Consider a 3d grayscale image with a total of $N = L \times L \times L$ voxels, binarized at a given threshold, and attribute "spins" $\sigma_i = -1$ (black) to voxels below the threshold, and spins $\sigma_i = +1$ (white) to voxels above the threshold. The threshold can assume any value between the observed minimum and maximum voxel values, where choice of the minimum for threshold yields a fully black image, choice of the maximum yields a white image, and intermediate threshold values yield images with varying concentrations of black and white voxels.

Each 1d word, 2d plaquette, or 3d block with $w$ voxels can be found in one of the $2^w$ possible configurations, each with a particular value of order parameter (equivalent of magnetization in statistical physics)

$$m_c = \sum_{i=1}^{w} \sigma_i^c \qquad , \qquad c = 1, \dots, 2^w \quad , \tag{1}$$

where $\sigma_i^c = \pm 1$ is the value of spin $i$ within configuration $c$. For 1d words of length $l$ there is a total of $N_l = 3L^2(L - l + 1)$ distinct samples of size $l$ in the original image ($L^2$ lines of length $l$, in three directions), for 2d plaquettes of linear size $l$ there are $N_l = 3L(L - l + 1)^2$ samples ($(L - l + 1)^2$ positions for plaquettes of size $l \times l$, for $L$ planes in each of the three directions), and for 3d blocks of linear size $l$ there are $N_l = (L - l + 1)^3$ samples (distinct positions for blocks of size $l \times l \times l$ ).

Let us now denote by $f_c = n_c/N_l$ the observed relative frequencies (probabilities) of the individual configurations in the dataset binarized at a given threshold, where $n_c$ is the observed number of occurrences of configuration $c = 1, \dots, 2^w$, satisfying $\sum_{c=1}^{2^w} n_c = N_l$. Following [1] let us also attribute to each configuration probability (Boltzmann factor)

$$p_c(T) = \frac{e^{-E_c/T}}{Z(T)} \tag{2}$$

as a function of a continuous disorder parameter $T$ (equivalent of temperature in statistical physics), where $E_c$ is energy of configuration $c$ and

$$Z(T) = \sum_{c=1}^{2^n} e^{-E_c/T} \tag{3}$$

is the partition function.

If the original relative frequencies are attributed $T = 1$ so that $f_c \equiv p_c(T = 1)$, as the sum of relative frequencies is by construction $\sum_{c=1}^{2^n} f_c = 1$ the set of energy values $E_c$ is defined by the observed relative frequencies up to an additive constant. Thus, the choice of $Z(T = 1)$ defines the energy scale (position of the ground state) and by choosing $Z(T = 1) = 1$ we have

$$E_c = -\ln f_c \ , \tag{4}$$

where the ground state corresponds to the configuration with the largest observed frequency. With this choice of scale, for $T = 1$ entropy is given by

$$S = -\sum_{c=1}^{2^n} f_c \ln f_c = \sum_{c=1}^{2^n} f_c E_c \equiv \langle E \rangle \ . \tag{5}$$

For the configuration probabilities for different $T$ values we can now write

$$p_c(T) = \frac{f_c^{\frac{1}{T}}}{Z(T)} \quad , \tag{6}$$

where the partition function is

$$Z(T) = \sum_{c=1}^{2^n} f_c^{\frac{1}{T}} \quad . \tag{7}$$

In the limit $T \to 0$ only the ground state configuration (the one with the largest observed relative frequency) assumes unit probability while probability of all the other configurations is zero, and in the limit $T \to \infty$ all the configurations become equiprobable.

Order parameter (magnetization) is now given by

$$\langle M(T) \rangle = \frac{1}{Z(T)} \sum_{c=1}^{2^n} m_c \, f_c^{\ 1/T} \ . \tag{8}$$

The zero field susceptibility per spin (normalized magnetization variance) is given by [2]

$$\chi(T) = \left. \frac{\partial M}{\partial H} \right|_{H=0} = \frac{\langle M(T)^2 \rangle - \langle M(T) \rangle^2}{nT} \quad , \tag{9}$$

where

$$\langle M(T)^2 \rangle = \frac{1}{Z(T)} \sum_{c=1}^{2^n} m_c^2 \, f_c^{\ 1/T} \quad . \tag{10}$$

Finally, specific heat per spin (normalized energy variance) is given by [1,2]

$$C(T) = \frac{1}{n} \frac{\partial U}{\partial T} = \frac{\langle E(T)^2 \rangle - \langle E(T) \rangle^2}{nT^2} \quad , \tag{11}$$

where

$$\langle E(T)\rangle = -\frac{1}{Z(T)}\sum_{c=1}^{2^n} \ln f_c \ f_c^{\frac{1}{T}} \quad , \tag{12}$$

and

$$\langle E(T)^2\rangle = \frac{1}{Z(T)}\sum_{c=1}^{2^n} (\ln f_c)^2 \ f_c^{\ 1/T} \quad , \tag{13}$$

Equations (8), (9) and (11) present the signatures of the order parameter (magnetization) and response functions (susceptibility and specific heat) of the relative frequencies (probabilities) $f_c$, $c = 1, \dots, 2^n$ encountered in the dataset for any given threshold.

It should be stressed here that the above formalism is quite general, it can be applied to study time series, two dimensional images or three-dimensional data.

## Results

### *3d X-ray soil sample image*

In what follows formulas (8), (9) and (11) are applied to samples extracted from the 790x790x790 grayscale soil image for different threshold values. In Tab. 1 the total number of samples is listed together with the total number of possible sample configurations for 1, 2 and 3 dimensions. While the ratio of the number of possible sample configurations and the total number of samples cannot be considered statistically significant for $l = 28$ and $l = 32$ in the one dimensional case, for liner size $l = 6$ for two dimensional samples, and for $l = 3$ for three dimensions, it will be shown that these cases still exhibit finite size scaling in agreement with the other sample sizes, and shall therefore be included in the rest of this study.

**Table 1.** The number of samples extracted from the 790x790x790 grayscale image for 1, 2 and 3 dimensions, together with the total number of possible sample configurations, and the ratio between these two.

| Linear size | # of samples | # of configurations | ratio |
|---|---|---|---|
| $l$ | 1d | $2^l$ | |
| 4 | 1473500100 | 16 | 92093756 |
| 8 | 1466010900 | 256 | 5726605 |
| 12 | 1458521700 | 4096 | 356084 |
| 16 | 1451032500 | 65536 | 22141 |
| 20 | 1443543300 | 1048576 | 1377 |
| 24 | 1436054100 | 16777216 | 85.6 |
| 28 | 1428564900 | 268435456 | 5.32 |
| 32 | 1421075700 | 4294967296 | 0.33 |
| $l$ | 2d | $2^{l^2}$ | |
| 2 | 1475374770 | 16 | 92210923 |
| 3 | 1471637280 | 512 | 2874292 |
| 4 | 1467904530 | 65536 | 22398 |
| 5 | 1464176520 | 33554432 | 43.6 |
| 6 | 1460453250 | 68719476736 | 0.021 |
| $l$ | 3d | $2^{l^3}$ | |
| 2 | 491169069 | 256 | 1918629 |
| 3 | 489303872 | 134217728 | 3.64 |

The main results of this manuscript are presented in Figs. 1-3 for order parameter (magnetization) and response functions (specific heat and susceptibility) for 1d, 2d and 3d samples, respectively. In all three cases it is seen that the finite size response functions appear to diverge, converging from right to left below the threshold of 1300 Hounsfield units (HU), reminiscent of the Ising model finite size scaling for periodic boundary conditions [9], while above 1300 HU the response functions approach thermodynamic limit from left to right, reminiscent of the Ising model finite size scaling for open boundaries [9,10]. The threshold values were initially scanned throughout a range of values between the observed voxel HU minimum and maximum, upon which the values 1100, 1300 and 1500 HU were identified as good representatives of the crossover behavior.

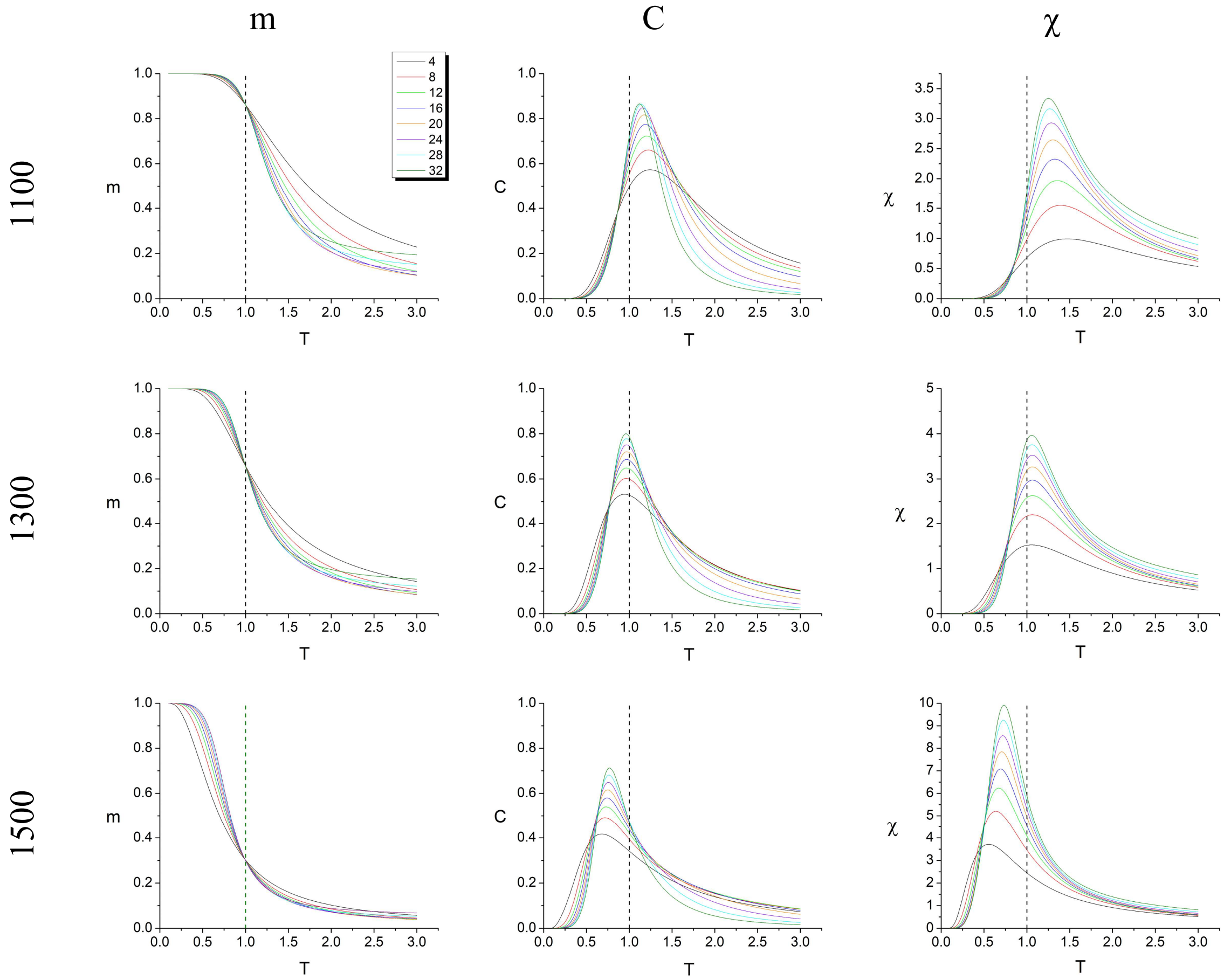

**Figure 1.** Magnetization $m$, specific heat $C$ and susceptibility $\chi$ for 1d samples of length 4-32 of the 3d soil image under study, for different threshold values of 1100-1500 HU.

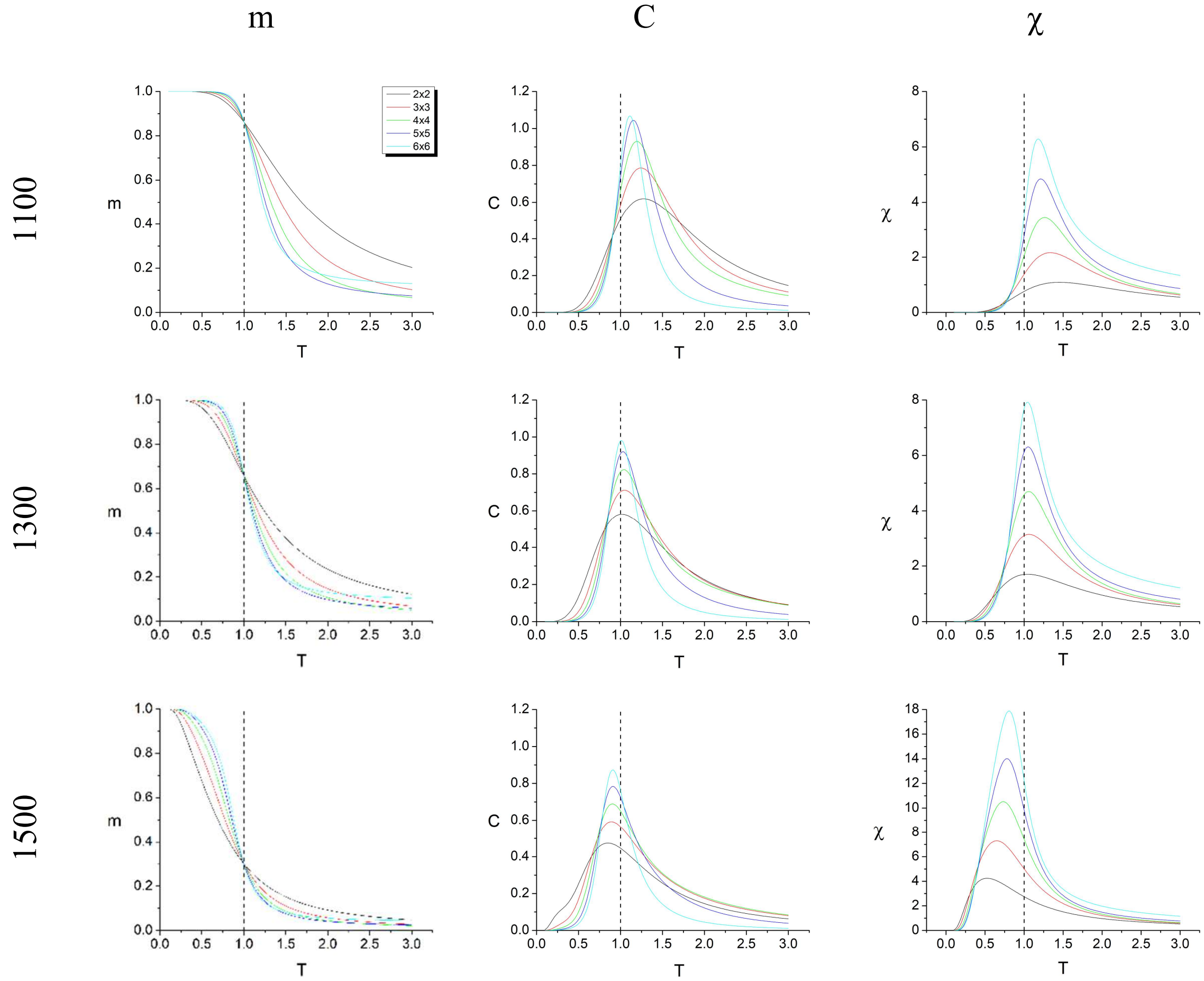

**Figure 2.** Magnetization $m$, specific heat $C$ and susceptibility $\chi$ for 2d samples of size 2x2 up to 6x6 of the 3d soil image under study, for different threshold values of 1100-1500 HU.

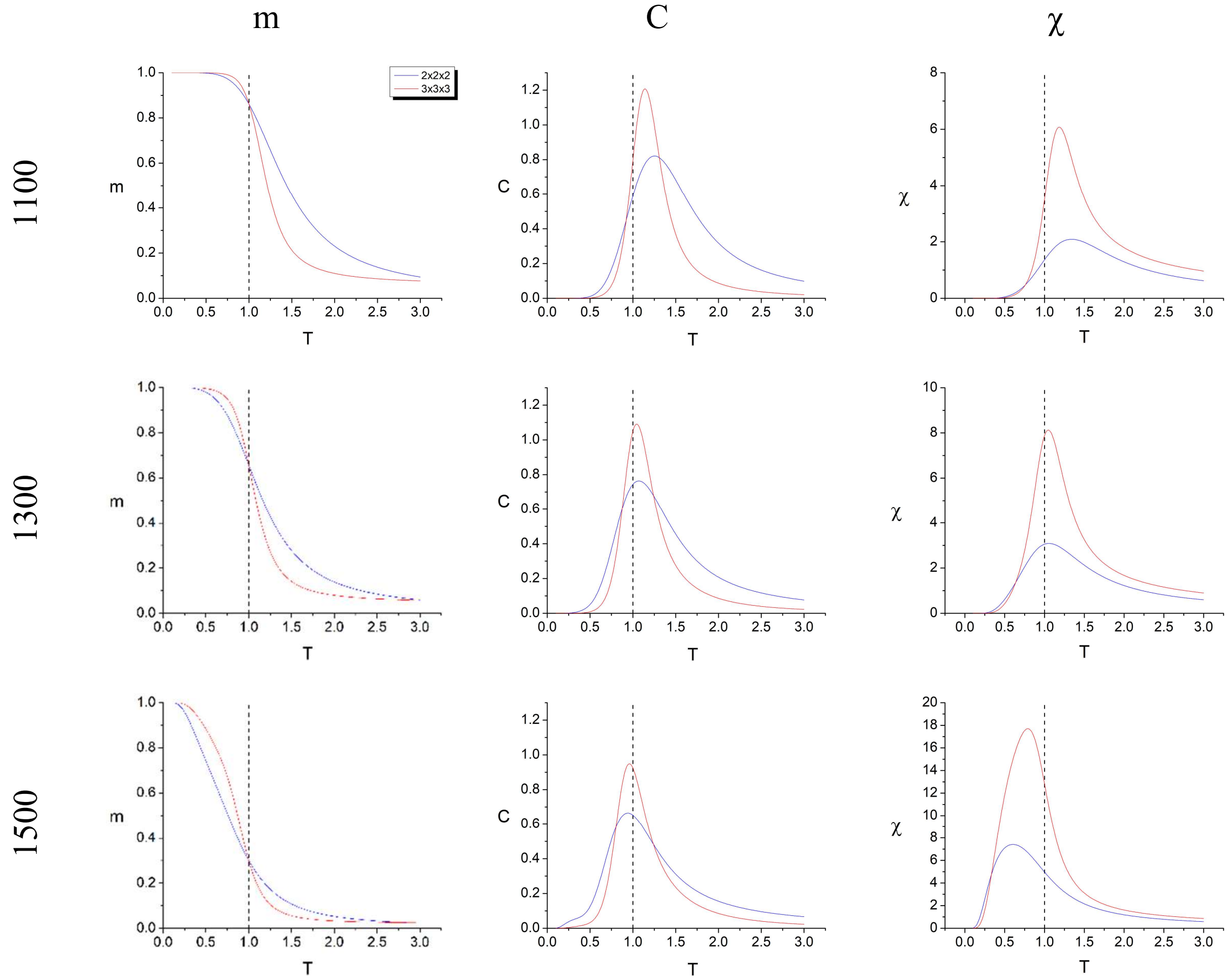

**Figure 3.** Magnetization $m$, specific heat $C$ and susceptibility $\chi$ for 3d samples of size 2x2x2 and to 3x3x3 of the 3d soil image under study, for different threshold values of 1100-1500 HU.

Critical behavior, characterized by divergence of response functions, is reached only in the thermodynamic limit (infinite system size), while for finite systems the response functions exhibit a finite peak that increases and approaches the critical temperature as the system size increases, which is the subject of finite size scaling theory. Close to the critical temperature $T_C$ curves of magnetization, specific heat and susceptibility collapse onto constant curves when plotted as $mL^{\beta/\nu}$, $CL^{-\alpha/\nu}$ and $\chi L^{-\gamma/\nu}$ versus $\tau L^{1/\nu}$ [2], where $L$ is the linear size of the system, $\tau = (T - T_C)/T_C$ is dimensionless temperature that measures distance from the critical point, $\nu$ is the critical exponent of the correlation length $\xi \sim |\tau|^{-\nu}$, and $\alpha$, $\beta$ and $\gamma$ are critical exponents of the specific heat, magnetization and susceptibility, respectively. Critical exponents are obtained by minimizing the distance between curves for different sizes, and here a Mote Carlo Markov Chain (MCMC) algorithm is employed to this end, implementing simulated annealing in the parameter space of the critical exponents α, β and γ. As the individual data points for different sample sizes do not coincide on the scaled temperature scale, they are binned (100 bins were used). The scaled $m$, $C$ and $\chi$ are then integrated inside each bin to find the RMS error for all finite size curves simultaneously, which is then minimized through an acceptance/rejection process, with a gradually adaptive acceptance criterion. The RMS decrease is always accepted, and RMS increase is accepted through an exponential criterion $e^{-\Delta_{RMS}/T_{MCMC}}$, where $\Delta_{RMS}$ is the RMS change produced by a tentative change of a critical exponent value, and $T_{MCMC}$ is the annealing "temperature" which starts with a high value in search of global RMS minima, and is gradually decreased (annealing) to improve the precision. MCMC simulation stops when the RMS does not decrease after 10000 consecutive trials. In particular, it turns out that the choice $T_C = 1.0$ (as also observed in [1] and [2]) and $\nu = 3.0$ yields the best finite size curve collapse across the spectrum of combinations of sampling choices, system sizes, and threshold values.In terms of finite size scaling of the Ising model, for periodic boundary conditions the response functions approach the thermodynamic limit from the right [9], while for open boundary conditions the response functions approach the thermodynamic limit from the left [9,10]. This is the consequence of the fact that for periodic boundary conditions the possible self-avoiding paths among any two sites (which determine the correlation length) are overestimated with respect to the infinite system (thermodynamic limit), while for open boundaries the possible paths between any two sites are comparatively underestimated. At the critical temperature the correlation length diverges together with the response functions, and the system becomes scale free. In the current case there are no paths among pairs of data points, but the correlation length can be over or underestimated for finite size samples. Collapsing the curves for finite size systems according to finite size scaling theory is shown in Figs. 4-6, and the critical exponents obtained through Monte Carlo Markov Chain for the best curve collapsing results are given in Tab. 2.

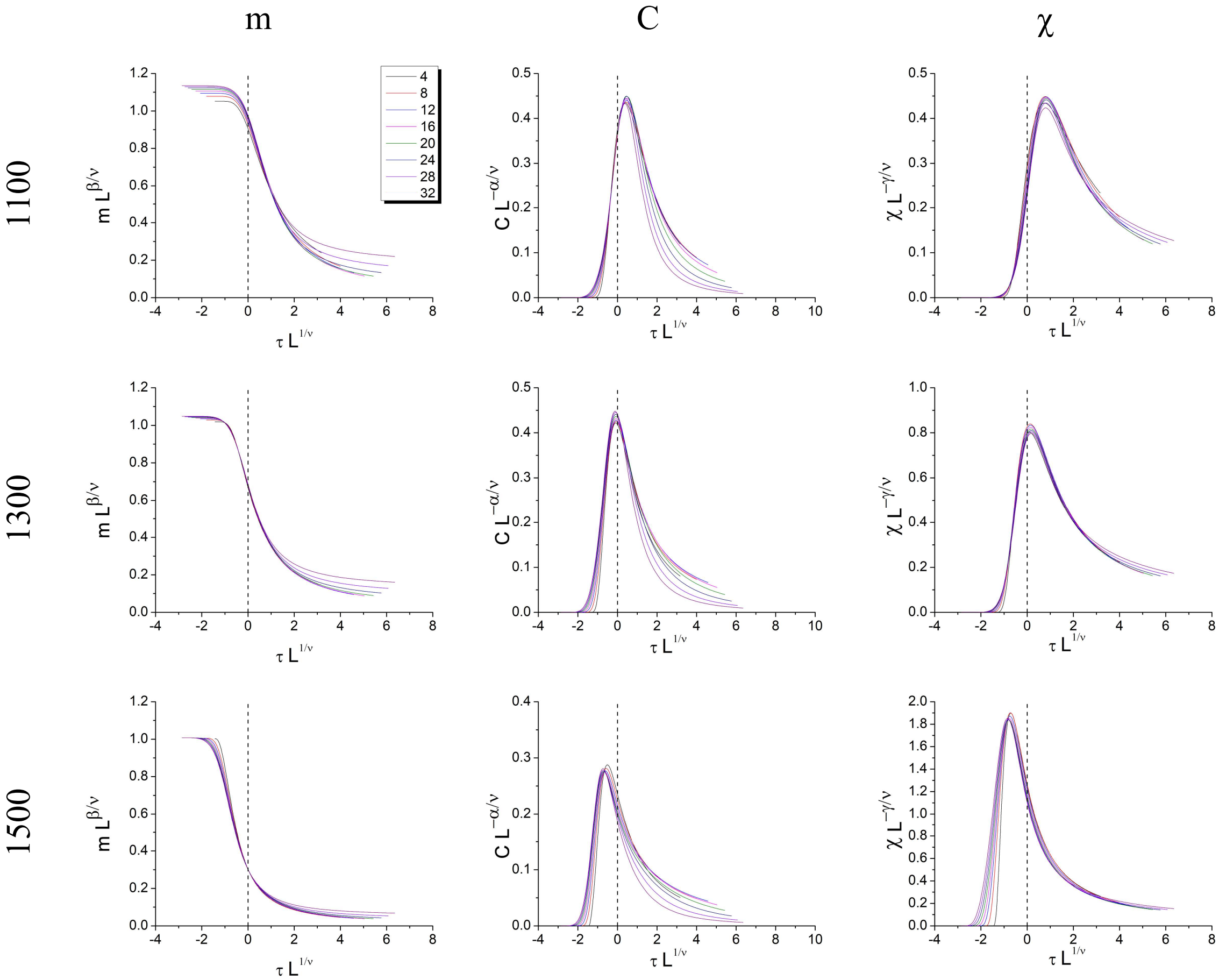

**Figure 4.** Finite size scaling of magnetization, specific heat and susceptibility for 1d samples.

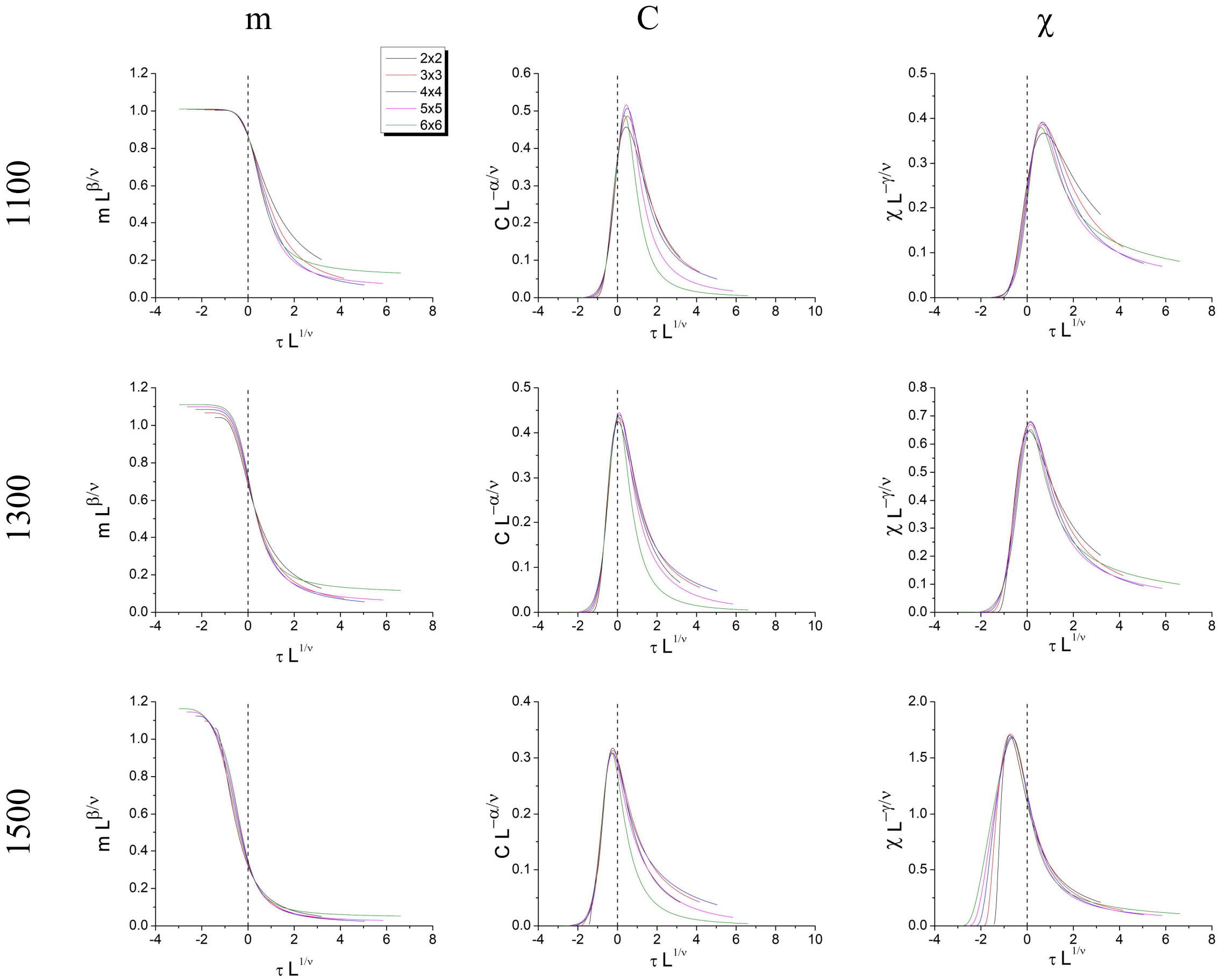

**Figure 5.** Finite size scaling of magnetization, specific heat and susceptibility for 2d samples.

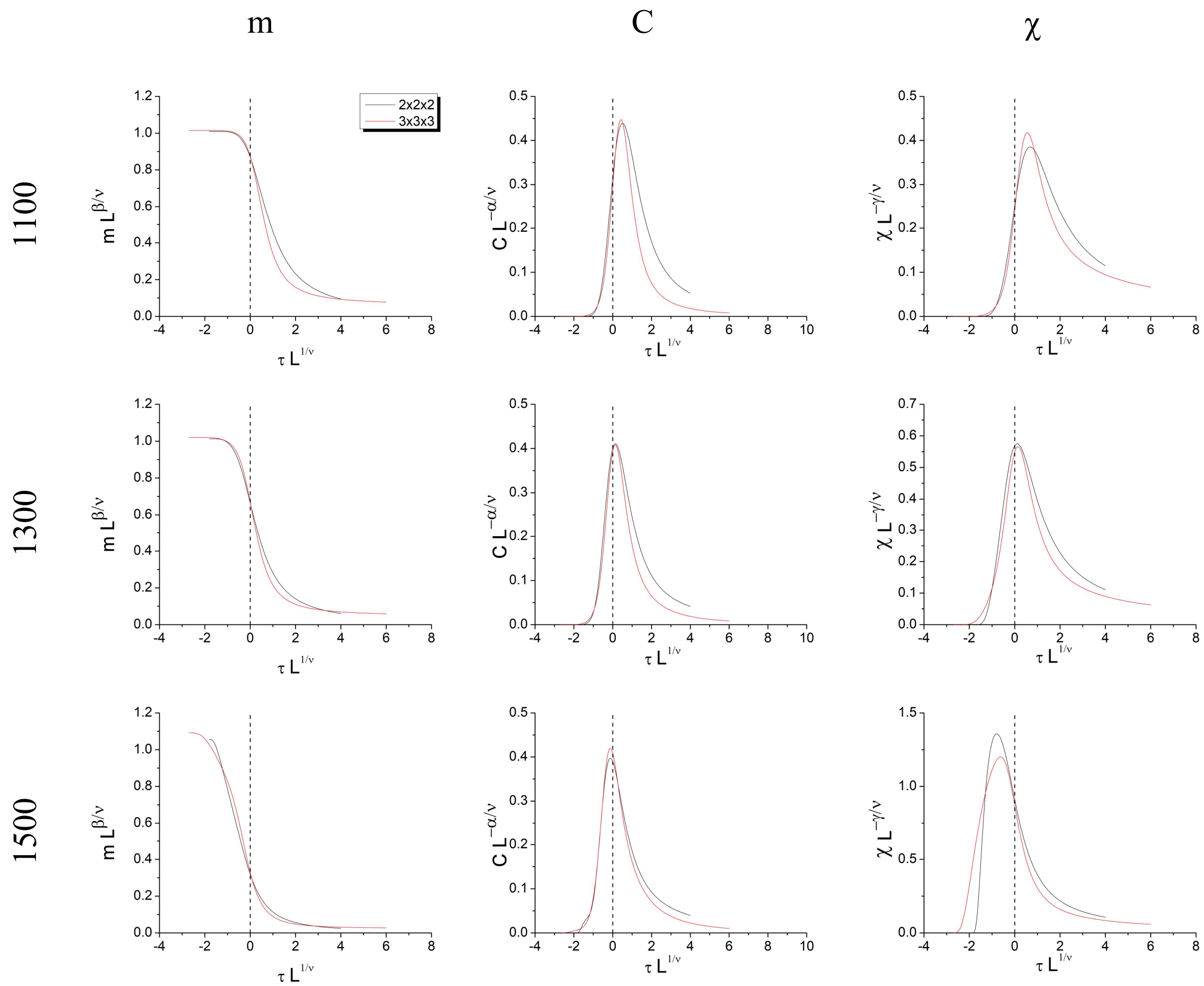

**Figure 6.** Finite size scaling of magnetization, specific heat and susceptibility for 3d samples.

**Table 2.** Critical exponents for $T_C = 1.0$ and $\nu = 3.0$

| | Threshold (HU) | $\alpha$ | $\beta$ | $\gamma$ |
|---|---|---|---|---|
| 1d | 1100 | 0.599 | 0.109 | 1.787 |
| | 1300 | 0.503 | 0.040 | 1.388 |
| | 1500 | 0.805 | 0.006 | 1.452 |
| 2d | 1100 | 0.656 | 0.008 | 2.349 |
| | 1300 | 0.677 | 0.088 | 2.091 |
| | 1500 | 0.869 | 0.126 | 1.979 |
| 3d | 1100 | 0.904 | 0.0130 | 2.436 |
| | 1300 | 0.895 | 0.0181 | 2.425 |
| | 1500 | 0.742 | 0.0794 | 2.449 |

These critical exponents presented in Tab. 2 do not seem to fall into any of the known universality classes, but the existence of novel "data universality classes" is yet to be explored.

In the current case the choice of threshold leads to over or underestimation of correlations, such that at ~1300 HU the system appears to be scale free: the response functions are centered at the critical temperature T=1 for different (increasing) sample sizes, for all the three sampling dimension choices. This finding in turn leads to the conjecture that connected voxel clusters found close to that threshold level should be scale free, and thus correspond to the REV in terms of soils structure. The 3d images of the soil sample with a threshold at 1100, 1300 and 1500 HU are shown in Fig. 7.

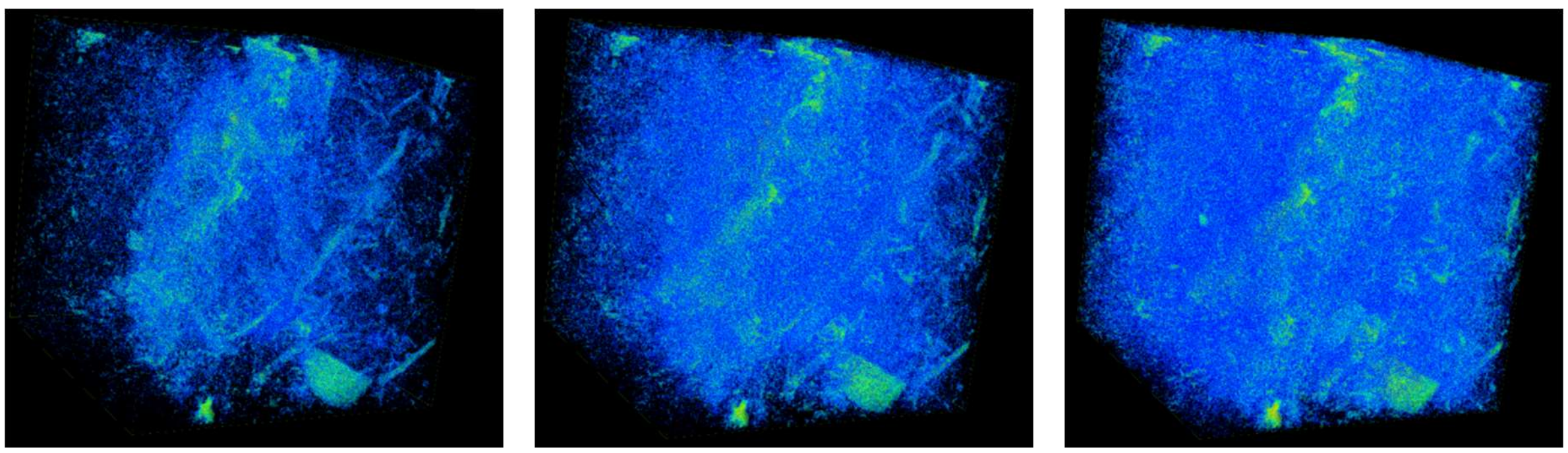

**Figure 7.** The 3d images of the soil sample with a threshold at 1100, 1300 and 1500 HU, from left to right.

Next, Hoshen-Kopelman analysis [11] was performed to examine the clusters in the vicinity of the transition threshold. A total of 11,439,660 clusters of connected voxels with radiodensity between 1250 and 1350 HU was identified, ranging in size from 1 to 1855 voxels. The cluster size distribution and the image with the 100 largest clusters is shown in Fig. 8., and the details of the dimensions of the 10 largest clusters are presented in Tab. 3.

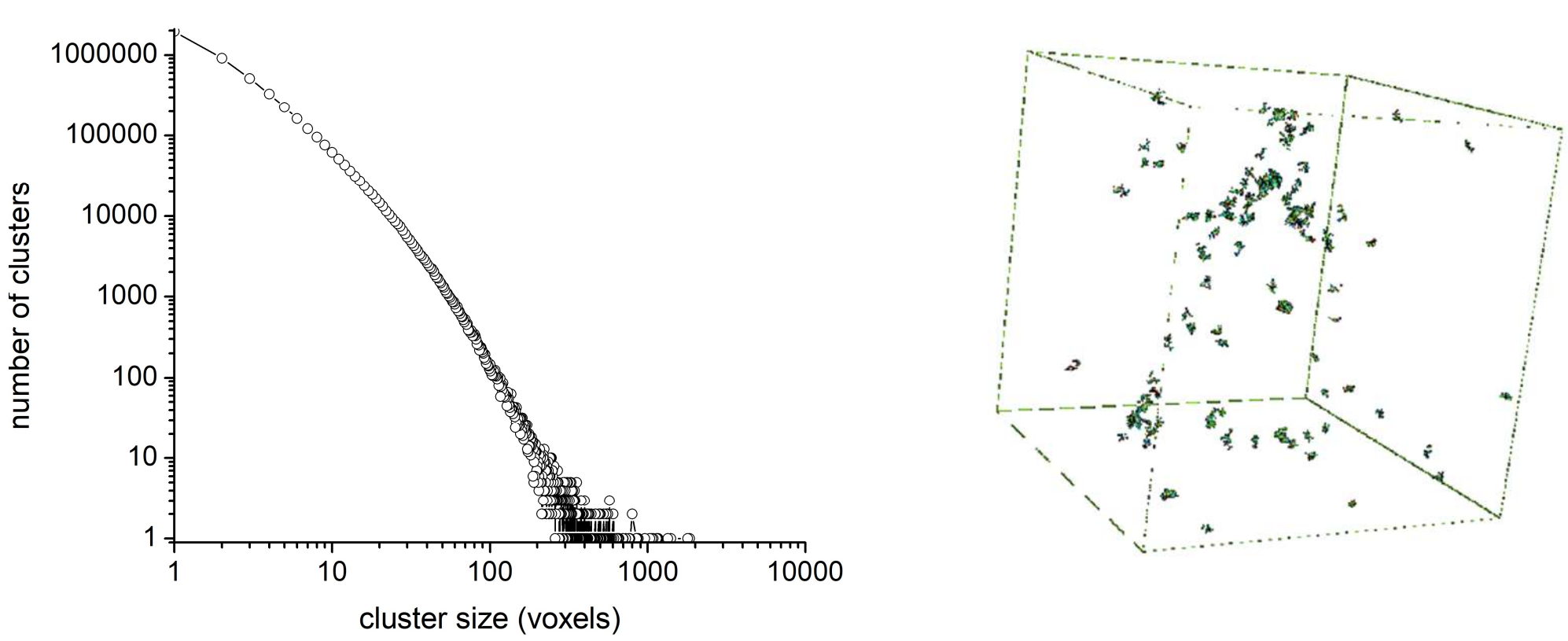

**Figure 8.** Cluster size distribution (left panel) ad the 100 largest clusters (right)

**Table 3.** Properties of the 10 largest clusters: size (in voxels), bounding box, volume of the bonding box, density (fraction of the bounding box occupied by cluster voxels), and the bounding box diagonal in voxels and millimeters.

| size | bounding box | volume | density | diagonal (voxels) | diagonal (mm) |
|---|---|---|---|---|---|
| 1855 | 36x28x27 | 27216 | 0.0682 | 30.080 | 1.35 |
| 1793 | 60x29x39 | 67860 | 0.0264 | 40.789 | 1.84 |
| 1405 | 30x36x26 | 28080 | 0.0500 | 30.395 | 1.37 |
| 1341 | 32x39x24 | 29952 | 0.0448 | 31.056 | 1.40 |
| 1165 | 26x41x28 | 29848 | 0.0390 | 31.020 | 1.40 |
| 1156 | 33x27x28 | 24948 | 0.0463 | 29.220 | 1.31 |
| 1154 | 23x31x30 | 21390 | 0.0540 | 27.759 | 1.25 |
| 1138 | 30x23x41 | 28290 | 0.0402 | 30.470 | 1.37 |
| 1085 | 29x25x33 | 23925 | 0.0454 | 28.815 | 1.30 |
| 1063 | 22x36x41 | 32472 | 0.0327 | 31.903 | 1.44 |

These phenomenological results suggest that a characteristic distance of approximately 30 voxels or 1.4 mm for the current 3d grayscale soil image may be considered to correspond to the "structural" Representative elementary volume. While the current soil sample is highly anisotropic and inhomogeneous as can be seen in Fig. 4, the overall structure at the threshold of ~1300 HU exhibits scale free behavior from the thermodynamic perspective of finite size scaling.

The implications of these phenomenological findings and their relevance for elucidating the phenomenon at hand are yet to be addressed by the soil science community.

## Conclusions

The current work demonstrates how the novel approach of "data thermodynamics" can shed new light on data in general, in one, two and three dimensions. The 3d soil sample used in this study was sampled for all three dimension choices, to check the comparative benefits. It is found that all three choices yield similar behavior - a threshold below which the response functions approach the thermodynamic limit from the right (reminiscent of the Ising model finite size scaling behavior for periodic boundary conditions [9]), and above which the response functions approach thermodynamic limit from left to right, reminiscent of the Ising model finite size scaling for open boundaries [9,10]. This observed threshold implies existence of a characteristic scale, where the response functions become scale free. Tests are currently under way using different datasets. While not all datasets exhibit the behavior observed here (right to left crossover of the increasing peaks of the response functions), preliminary results for the logarithmic returns of the Forex EUR/USD temporal series do, to be reported elsewhere.

The relevance of this characteristic scale strongly depends on the phenomenon at hand; while in the current case of an X-ray 3d soil sample it may be related to the evasive concept of Representative elementary volume, in the case of financial data, or neural brain activity it may help elucidate some previously unknown aspects of the phenomenon at hand. In summary, the formidable formalism of thermodynamics is not restricted to systems of interacting particles, rather, it can be generalized to study of any data, in 1, 2, or 3 (or perhaps more) dimensions. While the wider adoption of these concepts [1,2] over the last decade in different areas of knowledge appears to have been rather slow (or inexistent), it is the belief of the current author that far more research should be directed in this direction.

Finally, the critical exponents presented in Tab. 2 do not seem to fall into any of the known universality classes, but they may lead to novel "data universality classes", to help classify behavior of data in different fields of knowledge.

**Acknowledgments:** The author acknowledges support of Brazilian agency CNPq grant Nº 308782/2022-4, and the Brazilian agency CAPES through grant Nº 88887.937789/2024-00. Support of the X-ray Computed Tomography Laboratory and of the Multi-user Laboratory in Porous Media (LAMMEP) of the Nuclear Energy Department at UFPE is also acknowledged.